\documentclass[useAMS,usenatbib]{mn2e}
\usepackage[fleqn]{amsmath}
\usepackage{amssymb}
\usepackage{booktabs}
\usepackage{graphicx}
\newcommand{\fxf}{fxf@astro.ox.ac.uk}
\newcommand{\kevin}{kevins@astro.ox.ac.uk}

\title[Radial distribution of SNe Ia in early-type galaxies]{The
  radial distribution of type Ia supernovae in early-type galaxies:
  implications for progenitor scenarios}

\author[F.\ F\"orster, K.\ Schawinski]{Francisco
  F\"orster$^{1}$\thanks{\fxf} and
  Kevin Schawinski$^{1}$\thanks{\kevin}\smallskip \\$^{1}$Dept. of
  Physics, University of Oxford, Denys Wilkinson Building, Keble Road,
  Oxford, OX1 3RH, United Kingdom}

\begin{document}

% added comment here
\date{Accepted 2008 May 18.  Received 2008 May 12; in original form 2008 April 29.}

%% the date
%\date{\today}

%% maketitle
\maketitle

\label{firstpage}

%% abstract
%% --------------------------------------------------------------------------
\begin{abstract}

We study the radial distribution of supernova Ia (SNe Ia) in
morphologically selected early--type host galaxies from the
\emph{Sloan Digital Sky Survey} (SDSS) and discuss its implications
for the progenitor systems of SNe Ia. While new observations of
early-type galaxies suggest that they contain small fractions of young
stellar populations, they are also the most likely hosts for long time
delay SNe Ia. We find that there is no statistically significant
difference between the radial distribution of SNe Ia and the light
profile of their early--type host galaxies, which are dominated by
old, metal-rich stellar populations. This confirms the commonly
accepted idea that some SN Ia progenitors have time delays of the
order of several Gyr.

\end{abstract}

%% keywords
\begin{keywords}
supernovae: general, chemical: general.
\end{keywords}

%% ---------------------------------------------------------------------------------
\section{Introduction}

Type Ia supernovae (SNe Ia) have been extensively used as distance
indicators for cosmology \citep[see e.g.][]{rie98, per99} using the
\emph{empirical} Phillips relation \citep{phi93} and related
methods. They are also important for understanding galaxy evolution,
affecting the chemistry and energy budget of their hosts, and the star
formation and metallicity evolution of the Universe. Uncertainties in
both the progenitor scenarios and the explosion mechanism \citep[for a
review see][]{HN00} prevent us from having a convincing physical
picture that justifies their use as \emph{standardisable} candles at
high redshift and is able to reproduce their observed diversity in the
local Universe.

\subsection{The SN Ia time delay distribution} 

Linking SN Ia progenitor scenarios with cosmology, galaxy evolution
and the cosmic star formation history requires information on the time
between their formation and explosion, or time delay
distribution. Thermonuclear explosions seem to originate in white
dwarf stars (WDs), with main sequence (MS) lifetimes ranging from
$\sim 30$~Myr to several billion years, thus the minimum time delay
must be of the order of $\sim 30$~Myr. To narrow the constraints
details of possible \bf progenitor systems \rm need to be used.

Most proposed progenitor scenarios involve mass transfer on to a CO WD
that reaches the Chandrasekhar limit in a binary system. In these
scenarios the WD grows through either the expansion and Roche lobe
overflow of an evolved companion (\emph{single degenerate [SD]
scenarios}), or the slow release of gravitational waves, orbital
shrinking, Roche lobe overflow and merging of a compact double WD
system (\emph{double-degenerate [DD] scenarios}).

For the SD scenario it has been suggested that only systems that can
transfer enough mass under a critical rate can bring the WD star to
the Chandrasekhar limit \citep{nom91}. In binary population synthesis
(BPS) simulations most of the WDs that reach the Chandrasekhar mass
accrete matter from a slightly evolved MS star, the so called CO WD +
MS -- SD scenario \citep{HKN99, lan00, HP04}. In this channel the mass
of the companion determines both when accretion starts and the rate of
accretion. As a consequence, the time-delay distribution of this
channel seem to be relatively narrow, peaking at $\sim 670$~Myr and
rapidly becoming negligible after $\sim 1.5$~Gyr.

Several Gyr time--delay progenitors can also be produced in the SD
scenario when a relatively low mass red-giant (RG) star transfers
matter on to a CO WD star, in the CO WD + RG -- SD scenario
\citep{HKN96}. This scenario appears to be supported by observational
evidence \citep{2007Sci...317..924P}, although its relative
contribution is unclear \citep{HP04}.

The characteristic time delay in the DD scenario \citep{IT84, web84}
is the coalescence time--scale of the binary system, which depends
roughly on the fourth power of the separation of the double-degenerate
system \citep{sha83}. The time-delay distribution can be described by
a low time-delay cutoff ($\sim 30-100$~Myr), an approximately
power-law decline up to the age of the Universe and some secondary
peaks depending on the details of the BPS simulation. 

However, the expected accretion rates in the DD scenario are thought
to be too big to lead to successful thermonuclear explosions, with
most simulations leading to accretion-induced collapse (AIC) and the
formation of compact objects instead \citep[see][and references
therein]{SN98}. Rotation or more sophisticated neutrino physics may
give this channel a more significant role in the future
\citep{2007MNRAS.380..933Y}.

Observationally, a comparison between the cosmic supernova rate and
star formation history has led to indirect constraints on the time
delay distribution \citep[see e.g.][]{GYM04, str04,
2006ApJ...651..142H}, but uncertainties in both functional quantities
seem unlikely to give constraints with the required accuracy to
distinguish between different channels \citep{2006MNRAS.368.1893F,
2008arXiv0803.3793B}.

A better probe for clues regarding the SN Ia time delay distribution
might be to look at their individual host galaxies \citep[see
e.g.][]{man05, 2006ApJ...648..868S, 2008ApJ...673..999P} and their
galaxy environments \citep[see e.g.][]{2007ApJ...660.1165S}. With this approach uncertainties in the
star formation histories of individual galaxies can be better
controlled if accurate stellar population reconstructions are used.

\subsection{Early--type galaxies} 

The dominant stellar populations of early--type galaxies are old and
must have formed in the high redshift universe
\citep[see][and references therein]{2005ApJ...621..673T}.

However, the GALEX UV space telescope has led to the discovery that a
significant fraction of early--type galaxies in the low redshift
universe formed a few percent of their stellar mass in the last Gyr
\citep{2005ApJ...619L.111Y, 2006Natur.442..888S, 2007ApJS..173..619K,
2007ApJS..173..512S}. These episodes of residual star formation are
not apparent in the optical, but become prominent in the UV. This
opens up the possibility that the SNe Ia observed in apparently
passive early--type galaxies may be due to these small young
populations and so might have short time delays.

The bulk stellar population of massive early-type galaxies follow a
$r^{1/4}$ or de Vaucouleurs law \citep{1948AnAp...11..247D} and the
dynamical relaxation time--scale of these systems is much higher than
the age of the Universe \citep{1987gady.book.....B}.  High resolution
H$\beta$ maps have been used to trace young stellar populations in
early-type galaxies and they tend not to follow the typical de
Vaucouleurs profile of their hosts \citep{2006MNRAS.366.1151S,
2006MNRAS.369..497K}. Evidence for central molecular disks in
early-type galaxies supports this idea \citep{2007AJ....134.2148L,
2008MNRAS.tmp..490C}.

If the radial distribution of SNe Ia follows the light of the bulk
stellar population in early--type galaxies, then their progenitors
must have an age of at least the minimum between the relaxation
time--scale and the age of the bulk of the stellar population. We
attempt to quantitatively test whether the radial distribution of SNe
Ia follows a de Vaucouleurs law.

%% ---------------------------------------------------------------------------------

\section{Sample Selection} \label{sec:sample}

We create a sample of SN Ia early-type host galaxies by selecting all
SNe Ia in the CfA list of supernovae\footnote{See:
\texttt{http://www.cfa.harvard.edu/iau/lists/Supernovae.html}} whose
hosts have been covered by the Sloan Digital Sky Survey (SDSS,
\citealt{2000AJ....120.1579Y, 2002AJ....123..485S}). The SDSS is a
survey of half the Northern Sky, providing us with photometry in the
five optical filters \textit{u,g,r,i} and \textit{z}
\citep{1996AJ....111.1748F, 1998AJ....116.3040G}. We use the current
data release DR6 \citep{2008ApJS..175..297A}.

We visually inspect every SN Ia host galaxy observed by SDSS and
select only those with early-type morphology. Visual classification
avoids the introduction of biases associated with proxies for
morphology such as colour, concentration or structural parameters
\citep{2008arXiv0804.4483L}. From the remaining galaxies we select
those with half--light radii bigger than 2.4$\arcsec$, or six pixels
in the SDSS images. The final sample contained galaxies with a mean
and median half--light radius of 17.6$\arcsec$ and 12.9$\arcsec$, or
about 44 and 32 pixels, respectively. Their redshifts were in the
range 0.001 to 0.3, with a mean of 0.06 and median of 0.05.

About three quarters of the SNe in the final sample were dicovered by
the KAIT/LOSS SN survey, the SDSS SN survey and amateur astronomers in
almost equal proportions. The rest were discovered by various surveys
including the Nearby SN Factory, among others.

\section{Analysis} \label{sec:analysis}

We use elliptical projected radial coordinates and model the surface
brightness profiles of the early--type galaxies in our sample as
\citep{1968adga.book.....S}:
\begin{align}   \label{eq:sersic}
  I(r) \equiv I_{\rm e}~ \exp \biggl\lbrace - b_{\rm n} \biggl[
  \biggl(\frac{r}{r_{\rm e}}\biggr)^{1/n} - 1 \biggr] \biggr\rbrace,
\end{align}
where $n$ is known as the S\'ersic index (normally assumed to be four
in early--type galaxies) and by definition half of the light comes
from $r < r_{\rm e}$, with $I_{\rm e} \equiv I(r_{\rm e})$. These
requirements define $b_{\rm n}$, approximately $2 n - 1/3$.

Now, using the change of variable $u \equiv b_{\rm n}
\biggl(\frac{r}{r_{\rm e}}\biggr)^{1/n}$ one can show that the
probability density per unit variable $u$ of a photon coming at a
radius $r = r_{\rm e} (u / b_{\rm n})^n$ is:
\begin{align} \label{eq:fu}
  f_u(u) %= f_r(r) \frac{dr}{du} 
  = \frac{u^{2n - 1}
  e^{-u}}{\int_0^\infty t^{2 n - 1} e^{-t} dt} = \frac{u^{2 n -
  1}}{\Gamma(2 n)} e^{-u},
\end{align}
where $\Gamma(u) \equiv \int_0^\infty t^{u-1} e^{-t} dt$. This allows
one to define an equivalent SN Ia rate surface magnitude brightness:
\begin{align} \label{eq:muia}
  \mu_{\rm Ia} \equiv -2.5~ \ln \biggl(\frac{f_u(u)~ \Gamma(2n)}{u^{\rm
 2n - 1}}\biggr) = 2.5 ~ b_{\rm n}
 \biggl(\frac{r}{r_{\rm e}}\biggr)^{1/n},
\end{align}
which, for a given $n$, should have the same slope of the log of the
true surface brightness in $(r/r_{\rm e})^{1/n}$ space, but differ by
a constant.

Thus, the cumulative probability distribution per unit variable $u$
that a photon comes from inside a region of radius $r = r_{\rm e} (u /
b_{\rm n})^n$ is:
\begin{align} \label{eq:Fx}
  F_u(u) = P(2n,~u),
\end{align}
where $P(a,~u)$ is the regularised incomplete Gamma function defined
as $P(a,~u) \equiv \int_0^u t^{a - 1} e^{-t} dt~ /~ \Gamma(a)$.

With the distribution of normalised distances of observed SNe Ia it is
possible to test whether the SN Ia rate is proportional to the surface
brightness intensity in early--type galaxies.

\subsection{Systematic effects}

Since core collapse SNe are very rare in early-type hosts, SN type
misclassification is unlikely. However, early--type galaxies have
extended radial profiles that easily overlap with other galaxies in
the line of sight, making host galaxy identification problems
important at large radii. Moreover, several effects will occur at
small radii: different point--spread functions, deviations from the
S\'ersic profile close to galaxy cores, astrometric errors between
SDSS and SN discovery coordinates, or the Shaw effect
\citep{1979A&A....76..188S}, i.e. undetected SNe near the core of
galaxies due to the saturation of old photographic plates.  To
minimise these systematic errors we limit our sample to SNe with
intermediate radii.

SDSS positions are accurate to 0.1$\arcsec$
\citep{2003AJ....125.1559P} and only 15\% of the SNe in our sample
were discovered before 2000, which means that astrometric problems or
the Shaw effect are unlikely to be significant. We define a minimum SN
distance of 0.2 half--light radii, since this central zone is well
resolved in more than 80\% of the images. For the smallest allowed
half--light radii this corresponds to about five times the SDSS
astrometric accuracy, but about two times the SDSS resolution. We also
define an arbitrary maximum distance of four half-light radii, which
is the radius at which approximately 85\% of the light of the galaxy
is contained in a de Vaucouleurs profile. The resulting mean and
median SN distances from the galaxy cores were 18.2$\arcsec$ and
9.6$\arcsec$, about 45 and 24 pixels, or 13 and 7 times de resolution,
respectively.

Thus, we limit our sample to SNe with $0.2 < r / r_{\rm e} < 4$. When
more than one host candidate met these requirements we arbitrarily
chose the one closer to the SNe. This could bias our results to a
centrally concentrated distribution, but dim host galaxies that are
not seen in SDSS images could have the opposite effect.

\begin{figure}
\centering
\includegraphics[clip=true,width=0.8\hsize, angle=0, keepaspectratio]{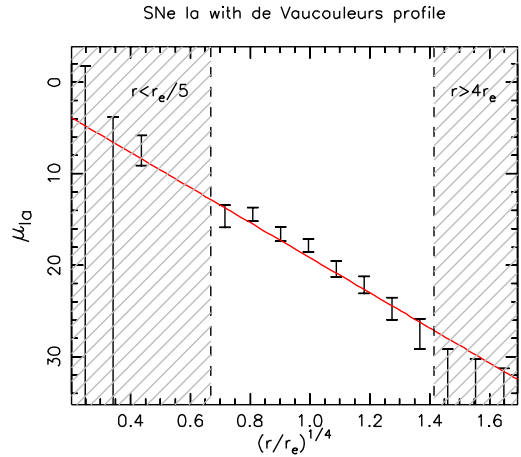}
\caption{Observed (error bars) values of $\mu_{\rm Ia}$, defined by
 eq.~(\ref{eq:muia}), vs the expected values (line) if SNe Ia follow a
 de Vaucouleurs law in early--type galaxies. The binning is robust at
 intermediate radii, with 5 or more elements in more than 80\% of the
 bins.}
\label{fig:SDSS_power}
\end{figure}

In Fig.~\ref{fig:SDSS_power} we plot the equivalent of the surface
brightness profile using type Ia SNe and assuming a de Vaucouleurs law
with the restricted normalised distances defined above. As a first
approximation it shows that SNe Ia follow their early--type host light
profiles. Other authors have studied the radial distribution of type
Ia SNe, but not using half--light radii normalised distances
\citep{2000ApJ...542..588I, 2007HiA....14..316B}. Recently,
\cite{2008arXiv0804.0909T} have also used half--light radii normalised
distances, but with photometrically typed SNe Ia at high redshift and
with a much lower relative resolution.

\subsection{Goodness of Fit test} 

\begin{figure*}
\centering
\vbox{
\includegraphics[width=0.4\hsize, angle=0, trim=0 0 0 0, keepaspectratio]{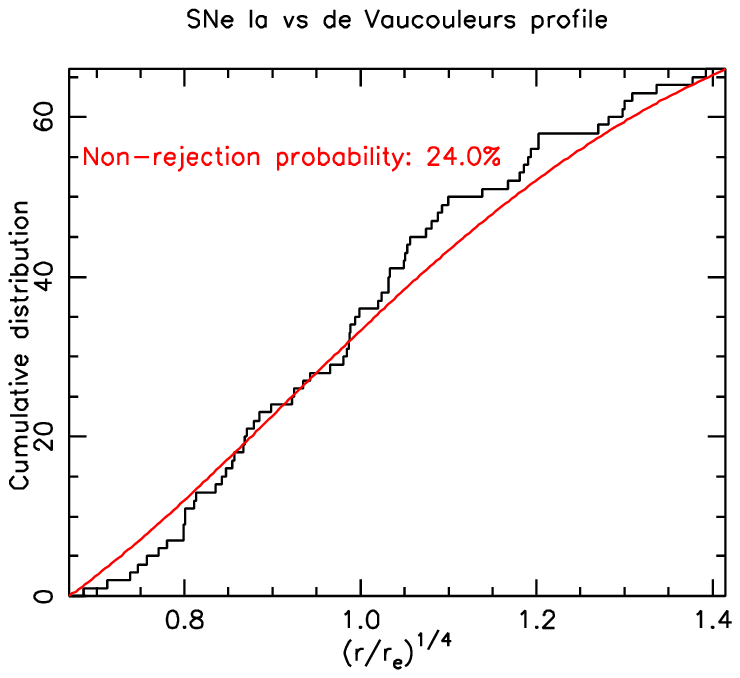}
\includegraphics[width=0.4\hsize, angle=0, trim=0 0 0 0,
  keepaspectratio]{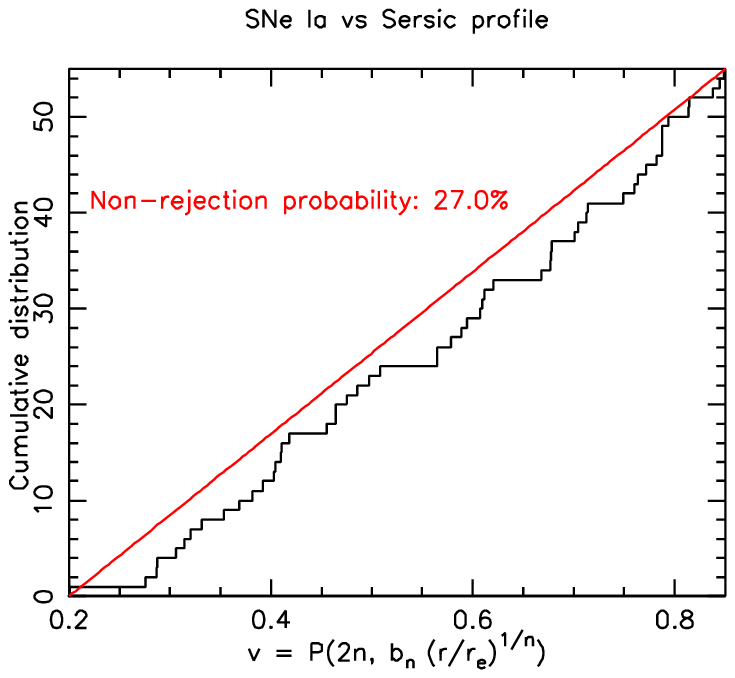}
}
\vbox{
\includegraphics[width=0.4\hsize, angle=0, trim=0 0 0 0, keepaspectratio]{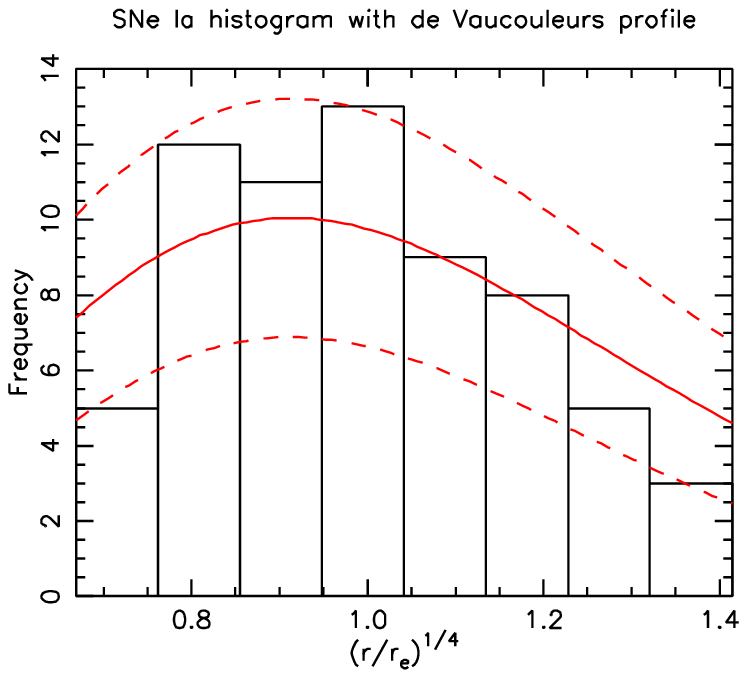}
\includegraphics[width=0.4\hsize, angle=0, trim=0 0 0 0,
  keepaspectratio]{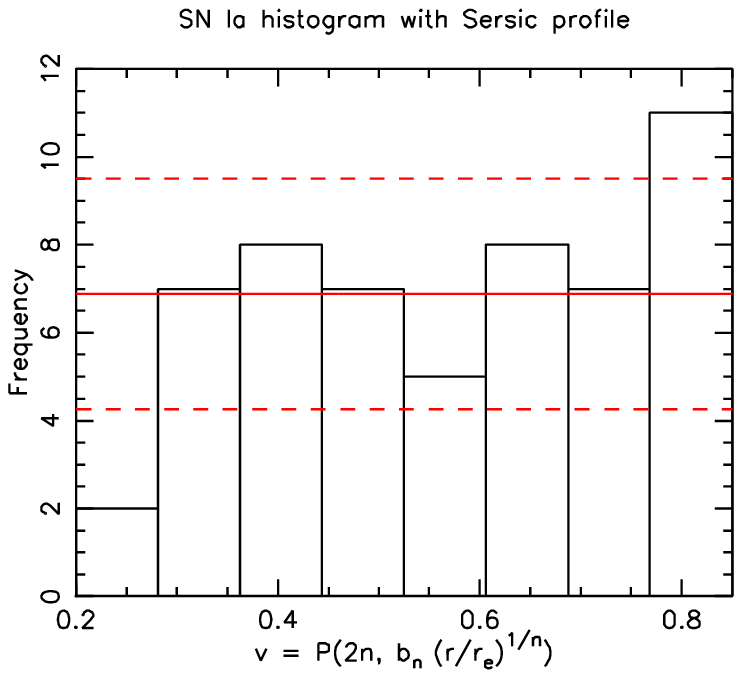}
}
\caption[KS test]{Kolmogorov Smirnov tests of the cumulative
  distribution of SNe Ia (top) and SN Ia histograms (bottom) assuming
  a de Vaucouleurs profile (left) or a S\'ersic profile (right) and
  the predicted distributions. The data was binned so that more than
  80\% of the bins would have at least 5 elements. Note that different
  selection criteria led to different sample sizes. }
\label{fig:test}
\end{figure*}

To quantitatively test the hypothesis that SNe Ia follow their
early--type hosts light profiles in this limited sample we normalise
the cumulative probability distribution in equation~(\ref{eq:Fx}), so
that $b_{\rm n}~ 0.2^{1/n} < u < b_{\rm n}~ 4^{1/n}$, and contrast it
with the observed cumulative distribution using a Kolmogorov-Smirnov
(KS) goodness of fit test. We conclude that for this restricted region
the SN Ia rate is statistically consistent with a de Vaucouleurs
profile (see Fig.~\ref{fig:test}).

However, it is known that early--type galaxies light profiles have a
continuous distribution of S\'ersic indices connecting dwarf to giant
ellipticals, increasing from approximately $n = 2$ to $n = 6$
\citep{GG03}. We have included a variable S\'ersic index in our
fitting procedure and found that the best--fitting S\'ersic indices
are distributed in approximately the same range, but have a slight
preference for lower values, with an average of 3.4 and a median of
3.3. This distribution might be a result of the way SNe Ia have
historically been discovered, which in many cases depends on
pre--selected galaxy samples.

To allow for this diversity to be included in our analysis we use a
new change of variable suggested by eq.~(\ref{eq:Fx}), namely $v \equiv
P(2n, u)$, or the fraction of the integrated light of the galaxy at
the position of the SNe if the profile was well described by
equation~(\ref{eq:sersic}). The new probability density and cumulative
probability density will be simply:
\begin{align}
  f_v(v) = 1 ~~~~~~~~{\rm and} ~~~~~~~~  F_v(v) &= v,
\end{align}
which are independent of $n$. This allows us to compare all galaxies
irrespective of their S\'ersic indices.

Hence, after finding the best--fitting values of $n$ for each
individual galaxy we set a cut for the SN sample based on the new
variable $v$ rather than $u$. Given that the half--light radii
resulting from the S\'ersic profile fitting were on average smaller,
we limit our sample to SNe within $0.2 < u < 0.85$, equivalent to $0.3
< r / r_{\rm e} < 4$ for the average S\'ersic index. We also select
galaxies were the best-fitting S\'ersic indices were in the expected
range $2 < n < 6$. The distribution of the variable $v$ contrasted to
$F_v(v)$ is in Fig.~\ref{fig:test}.

%% --------------------------------------------------------------------------------
\section{Conclusions and discussion} \label{sec:conclusions}

We have found that there is no statistically significant difference
between the radial dependence of the SN Ia rate in morphologically
selected early--type galaxies and their host surface brightness
profiles at intermediate radii. The probabilities of not rejecting the
hypotheses that SNe Ia follow a de Vaucouleurs or S\'ersic profile is
24.0\% and 27.0\%.

In this work we use host half--light radii ($r_{\rm e}$) normalised
distances, obtained after fitting de Vaucouleurs or S\'ersic profiles
on $r$ band SDSS images, and quantified any deviations from a de
Vaucouleurs or S\'ersic profile at intermediate radii. Intermediate
radii are defined in Section \ref{sec:analysis}.

%% --------------------------------------------------------------------------------
The main implications of this work is that the time delay of type Ia
SNe in early--type galaxies should be of the order of the age of their
host galaxies, or several Gyr. This assumes that young stellar
populations of early--type galaxies do not follow the light
distribution of their hosts, which is supported by recent observations
of central molecular disks and high--resolution SAURON H$\beta$ maps
and stellar populations reconstructions, showing that when young
($\lesssim$1 Gyr) stellar populations are present, they are mainly in
the central regions of their hosts. Even after including small radii
in our sample we did not see an excess of the SN Ia rate in the
central parts of early--type galaxies, but given all the possible
systematic errors of our method we cannot conclude if there is any
real deficit.

Our result does not exclude either the SD or DD scenarios in a picture of
multiple progenitor scenarios, but confirms the idea that very young
SN Ia progenitor scenarios are probably not dominant in early--type
galaxies. They rather indicate that long time delays of several Gyrs
for SNe Ia in early--type galaxies are dominant.

Interestingly, early-type galaxies are known to exhibit radial
gradients in metallicity, but not in age or $\alpha$-enhancement
\citep{2000A&A...360..911S, 2003A&A...407..423M,
2005ApJ...622..244W}. The metallicity gradients range all the way to
the outer parts of early-type galaxies, beyond two effective radii
\citep{2005ApJ...627..767M}. Given this, our result also suggests that
within the metallicity range exhibited by massive early-type galaxies,
the SN Ia rate appears not to be strongly affected by metallicity,
which hints that long time--delay SN Ia progenitor scenarios may not
have a very strong metallicity dependence. Note that a significant
dependence has been predicted for metallicities below solar
\citep[see][]{kob98, 2008arXiv0802.2471M}.

If the metallicity gradients and surface brightness profiles are
relics of the process of galaxy formation in early--type galaxies, our
result would suggest that the progenitors of long time--delay SN Ia
are as old as their host galaxies. 

Future studies of the radial distribution of sub--classes of SNe Ia,
e.g. SN1991bg and SN1991T--like events, should provide further
insights into whether there are multiple progenitor scenarios
\citep[see e.g.][]{2005ApJ...629L..85S}.

\section*{Acknowledgements}

We thank an anonymous referee for relevant comments that significantly
improved this manuscript. We are indebted to professional and amateur
astronomers that have made their SN discoveries public. We also thank
James Binney, Martin Bureau, Alison Crocker, Stephen Justham, Sadegh
Khochfar, Phillip Podsiadlowski, Katrien Steenbrugge, Mark Sullivan
and Christian Wolf for useful discussions, and Andr\'es Jordan for
providing the package PDL::Minuit. F.F. was supported by a Fundaci\'on
Andes -- PPARC Gemini studentship. K.S. was supported by the Henry
Skynner Junior Research Fellowship at Balliol College, Oxford.  Parts
of the analysis presented here made use of the Perl Data
Language\footnote{The Perl Data Language (PDL) has been developed by
K. Glazebrook, J. Brinchmann, J. Cerney, C. DeForest, D. Hunt,
T. Jenness, T. Luka, R. Schwebel, and C. Soeller and can be obtained
from http://pdl.perl.org}. The SN types and coordinates were obtained
from the Central Bureau for Astronomical Telegrams (CBAT) at the
Harvard-Smithsonian Center for
Astrophysics\footnote{\texttt{http://www.cfa.harvard.edu/iau/lists/Supernovae.html}}.
This publication makes use of data from the Sloan Digital Sky
Survey. The full acknowledgements can be found at
http://www.sdss.org/collaboration/credits.html.  We acknowledge the
use of NASA's SkyView facility http://skyview.gsfc.nasa.gov. This work
was supported in part through a European Research \& Training Network
on Type Ia Supernovae (HPRN-CT-20002-00303).

% the bibliography

\end{document}